\documentclass[showpacs,prd,10pt,twocolumn,superscriptaddress,aps,nofootinbib]{revtex4-1}

\usepackage{ae}
\usepackage{bm} 
\usepackage[dvipsnames]{xcolor}
\usepackage{amsmath}
\usepackage{amssymb}
\usepackage{amsfonts}
\usepackage{graphicx}
\usepackage{slashed}
\usepackage{wasysym}
\usepackage{units}
\usepackage{ulem}
\usepackage{upgreek}
\usepackage{chemformula}
\usepackage{siunitx}[=v2]
\usepackage{soul}
\usepackage{hyperref}

\newcommand{\Eqref}[1]{Eq.~\eqref{#1}}

\allowdisplaybreaks

\newcommand{\um}[0]{$\upmu$m}

\begin{document}

\setlength{\unitlength}{1mm}

\title{The Darkfield Approach to Measuring Vacuum Birefringence and Light-by-Light Couplings \\ A Proof-of-Principle Experiment}

\author{Michal~\v{S}m\'{\i}d}\email{m.smid@hzdr.de}
\affiliation{Helmholtz-Zentrum Dresden-Rossendorf, Bautzner Landstra\ss e 400, 01328 Dresden, Germany}

\author{Pooyan~Khademi}\email{pooyan.khademi@uni-jena.de}
\affiliation{Helmholtz-Institut Jena, Fr\"obelstieg 3, 07743 Jena, Germany}
\affiliation{GSI Helmholtzzentrum f\"ur Schwerionenforschung, Planckstra\ss e 1, 64291 Darmstadt, Germany}
\affiliation{Faculty of Physics and Astronomy, Friedrich-Schiller-Universit\"at Jena, 07743 Jena, Germany}

\author{Michal~Andrzejewski}
\affiliation{European XFEL, Holzkoppel 4, 22869 Schenefeld, Germany}

\author{Carsten~B\"ahtz}
\affiliation{Helmholtz-Zentrum Dresden-Rossendorf, Bautzner Landstra\ss e 400, 01328 Dresden, Germany}

\author{Erik~Brambrink}
\affiliation{European XFEL, Holzkoppel 4, 22869 Schenefeld, Germany}

\author{Tomáš~Burian}
\affiliation{Department of Radiation and Chemical Physics, Institute of Physics, Czech Academy of Sciences, Na Slovance 2, 182 00 Prague 8, Czech Republic}

\author{Jakub~Bulička}
\affiliation{Department of Radiation and Chemical Physics, Institute of Physics, Czech Academy of Sciences, Na Slovance 2, 182 00 Prague 8, Czech Republic}
\affiliation{Department of Surface and Plasma Science, Faculty of Mathematics and Physics, Charles University in Prague, V Holešovičkách 2, 182 00 Prague 8, Czech Republic}

\author{Samuele~Di~Dio~Cafiso}
\affiliation{Helmholtz-Zentrum Dresden-Rossendorf, Bautzner Landstra\ss e 400, 01328 Dresden, Germany}

\author{Jindřich~Chalupský}
\affiliation{Department of Radiation and Chemical Physics, Institute of Physics, Czech Academy of Sciences, Na Slovance 2, 182 00 Prague 8, Czech Republic}

\author{Thomas~E.~Cowan}
\affiliation{Helmholtz-Zentrum Dresden-Rossendorf, Bautzner Landstra\ss e 400, 01328 Dresden, Germany}

\author{Sebastian~G\"ode}
\affiliation{European XFEL, Holzkoppel 4, 22869 Schenefeld, Germany}

\author{Jörg~Grenzer}
\affiliation{Helmholtz-Zentrum Dresden-Rossendorf, Bautzner Landstra\ss e 400, 01328 Dresden, Germany}

\author{Věra~Hájková}
\affiliation{Department of Radiation and Chemical Physics, Institute of Physics, Czech Academy of Sciences, Na Slovance 2, 182 00 Prague 8, Czech Republic}

\author{Peter~Hilz}
\affiliation{Helmholtz-Institut Jena, Fr\"obelstieg 3, 07743 Jena, Germany}
\affiliation{GSI Helmholtzzentrum f\"ur Schwerionenforschung, Planckstra\ss e 1, 64291 Darmstadt, Germany}
\affiliation{Faculty of Physics and Astronomy, Friedrich-Schiller-Universit\"at Jena, 07743 Jena, Germany}

\author{Willi~Hippler}
\affiliation{Helmholtz-Institut Jena, Fr\"obelstieg 3, 07743 Jena, Germany}
\affiliation{GSI Helmholtzzentrum f\"ur Schwerionenforschung, Planckstra\ss e 1, 64291 Darmstadt, Germany}
\affiliation{Faculty of Physics and Astronomy, Friedrich-Schiller-Universit\"at Jena, 07743 Jena, Germany}

\author{Hauke~H\"oppner}
\affiliation{Helmholtz-Zentrum Dresden-Rossendorf, Bautzner Landstra\ss e 400, 01328 Dresden, Germany}

\author{Alžběta~Horynová}
\affiliation{Department of Radiation and Chemical Physics, Institute of Physics, Czech Academy of Sciences, Na Slovance 2, 182 00 Prague 8, Czech Republic}
\affiliation{Department of Nuclear Chemistry, Faculty of Nuclear Science and Physical Engineering, Czech Technical University in Prague, Břehová 7, 115 19 Prague 1, Czech Republic}

\author{Lingen~Huang}
\affiliation{Helmholtz-Zentrum Dresden-Rossendorf, Bautzner Landstra\ss e 400, 01328 Dresden, Germany}

\author{Oliver~Humphries}
\affiliation{European XFEL, Holzkoppel 4, 22869 Schenefeld, Germany}

\author{Šimon~Jelínek}
\affiliation{Department of Radiation and Chemical Physics, Institute of Physics, Czech Academy of Sciences, Na Slovance 2, 182 00 Prague 8, Czech Republic}
\affiliation{Laser Plasma Department, Institute of Plasma Physics, Czech Academy of Sciences, Za Slovankou 3, 182 00 Prague 8, Czech Republic}
\affiliation{Department of Surface and Plasma Science, Faculty of Mathematics and Physics, Charles University in Prague, V Holešovičkách 2, 182 00 Prague 8, Czech Republic}

\author{Libor~Juha}
\affiliation{Department of Radiation and Chemical Physics, Institute of Physics, Czech Academy of Sciences, Na Slovance 2, 182 00 Prague 8, Czech Republic}

\author{Felix~Karbstein}\email{f.karbstein@hi-jena.gsi.de}
\affiliation{Helmholtz-Institut Jena, Fr\"obelstieg 3, 07743 Jena, Germany}
\affiliation{GSI Helmholtzzentrum f\"ur Schwerionenforschung, Planckstra\ss e 1, 64291 Darmstadt, Germany}
\affiliation{Faculty of Physics and Astronomy, Friedrich-Schiller-Universit\"at Jena, 07743 Jena, Germany}

\author{Alejandro~Laso-Garcia}
\affiliation{Helmholtz-Zentrum Dresden-Rossendorf, Bautzner Landstra\ss e 400, 01328 Dresden, Germany}

\author{Robert~L\"otzsch}
\affiliation{Helmholtz-Institut Jena, Fr\"obelstieg 3, 07743 Jena, Germany}
\affiliation{GSI Helmholtzzentrum f\"ur Schwerionenforschung, Planckstra\ss e 1, 64291 Darmstadt, Germany}
\affiliation{Faculty of Physics and Astronomy, Friedrich-Schiller-Universit\"at Jena, 07743 Jena, Germany}

\author{Aim\'e~Matheron}
\affiliation{Helmholtz-Institut Jena, Fr\"obelstieg 3, 07743 Jena, Germany}
\affiliation{GSI Helmholtzzentrum f\"ur Schwerionenforschung, Planckstra\ss e 1, 64291 Darmstadt, Germany}
\affiliation{Faculty of Physics and Astronomy, Friedrich-Schiller-Universit\"at Jena, 07743 Jena, Germany}

\author{Masruri~Masruri}
\affiliation{European XFEL, Holzkoppel 4, 22869 Schenefeld, Germany}

\author{Motoaki~Nakatsutsumi}
\affiliation{European XFEL, Holzkoppel 4, 22869 Schenefeld, Germany}

\author{Alexander~Pelka}
\affiliation{European XFEL, Holzkoppel 4, 22869 Schenefeld, Germany}

\author{Gerhard~G.~Paulus}
\affiliation{Helmholtz-Institut Jena, Fr\"obelstieg 3, 07743 Jena, Germany}
\affiliation{GSI Helmholtzzentrum f\"ur Schwerionenforschung, Planckstra\ss e 1, 64291 Darmstadt, Germany}
\affiliation{Faculty of Physics and Astronomy, Friedrich-Schiller-Universit\"at Jena, 07743 Jena, Germany}

\author{Thomas~R.~Preston}
\affiliation{European XFEL, Holzkoppel 4, 22869 Schenefeld, Germany}

\author{Lisa~Randolph}
\affiliation{European XFEL, Holzkoppel 4, 22869 Schenefeld, Germany}

\author{Alexander~S\"avert}
\affiliation{Helmholtz-Institut Jena, Fr\"obelstieg 3, 07743 Jena, Germany}
\affiliation{GSI Helmholtzzentrum f\"ur Schwerionenforschung, Planckstra\ss e 1, 64291 Darmstadt, Germany}
\affiliation{Faculty of Physics and Astronomy, Friedrich-Schiller-Universit\"at Jena, 07743 Jena, Germany}

\author{Hans-Peter~Schlenvoigt}
\affiliation{Helmholtz-Zentrum Dresden-Rossendorf, Bautzner Landstra\ss e 400, 01328 Dresden, Germany}

\author{Jan~Patrick~Schwinkendorf}
\affiliation{Helmholtz-Zentrum Dresden-Rossendorf, Bautzner Landstra\ss e 400, 01328 Dresden, Germany}
\affiliation{European XFEL, Holzkoppel 4, 22869 Schenefeld, Germany}

\author{Thomas~St\"ohlker}
\affiliation{Helmholtz-Institut Jena, Fr\"obelstieg 3, 07743 Jena, Germany}
\affiliation{GSI Helmholtzzentrum f\"ur Schwerionenforschung, Planckstra\ss e 1, 64291 Darmstadt, Germany}
\affiliation{Faculty of Physics and Astronomy, Friedrich-Schiller-Universit\"at Jena, 07743 Jena, Germany}

\author{Toma~Toncian}
\affiliation{Helmholtz-Zentrum Dresden-Rossendorf, Bautzner Landstra\ss e 400, 01328 Dresden, Germany}

\author{Monika~Toncian}
\affiliation{Helmholtz-Zentrum Dresden-Rossendorf, Bautzner Landstra\ss e 400, 01328 Dresden, Germany}

\author{Maxim~Valialshchikov}
\affiliation{Helmholtz-Institut Jena, Fr\"obelstieg 3, 07743 Jena, Germany}
\affiliation{GSI Helmholtzzentrum f\"ur Schwerionenforschung, Planckstra\ss e 1, 64291 Darmstadt, Germany}
\affiliation{Faculty of Physics and Astronomy, Friedrich-Schiller-Universit\"at Jena, 07743 Jena, Germany}

\author{Sripati~V.~Rahul}
\affiliation{European XFEL, Holzkoppel 4, 22869 Schenefeld, Germany}

\author{Vojtěch~Vozda}
\affiliation{Department of Radiation and Chemical Physics, Institute of Physics, Czech Academy of Sciences, Na Slovance 2, 182 00 Prague 8, Czech Republic}

\author{Edgar~Weckert}
\affiliation{Deutsches Elektronen-Synchrotron DESY, Notkestrasse 85, 22607 Hamburg, Germany}

\author{Colin~Wessel}
\affiliation{Helmholtz-Institut Jena, Fr\"obelstieg 3, 07743 Jena, Germany}
\affiliation{GSI Helmholtzzentrum f\"ur Schwerionenforschung, Planckstra\ss e 1, 64291 Darmstadt, Germany}
\affiliation{Faculty of Physics and Astronomy, Friedrich-Schiller-Universit\"at Jena, 07743 Jena, Germany}

\author{Jan Wild}
\affiliation{Department of Surface and Plasma Science, Faculty of Mathematics and Physics, Charles University in Prague, V Holešovičkách 2, 182 00 Prague 8, Czech Republic}

\author{Ulf~Zastrau}
\affiliation{European XFEL, Holzkoppel 4, 22869 Schenefeld, Germany}

\author{Matt~Zepf}\email{m.zepf@hi-jena.gsi.de}
\affiliation{Helmholtz-Institut Jena, Fr\"obelstieg 3, 07743 Jena, Germany}
\affiliation{GSI Helmholtzzentrum f\"ur Schwerionenforschung, Planckstra\ss e 1, 64291 Darmstadt, Germany}
\affiliation{Faculty of Physics and Astronomy, Friedrich-Schiller-Universit\"at Jena, 07743 Jena, Germany}

\date{\today}

\begin{abstract}
 Vacuum fluctuations give rise to effective nonlinear interactions between electromagnetic fields. These generically modify the characteristics of light traversing a strong-field region. 
 X-ray free-electron lasers (XFELs) constitute a particularly promising probe, due to their brilliance, the possibility of precise control and favourable frequency scaling.
 However, the nonlinear vacuum response is very small even when probing a tightly focused high-intensity laser field with XFEL radiation and direct measurement of light-by-light scattering of real photons and the associated fundamental physics constants of the quantum vacuum has not been possible to date.
 Achieving a sufficiently good signal-to-background separation is key to a successful quantum vacuum experiment.
 To master this challenge, a darkfield detection concept has recently been proposed.
 Here we present the results of a proof-of-principle experiment validating this approach at the High Energy Density scientific instrument of the European X-Ray Free Electron Laser.

\end{abstract}

\maketitle

\section{Introduction}

The main challenge of experiments aiming at measuring the nonlinear response of the quantum vacuum to strong macroscopic electromagnetic fields is to discriminate the small signal component from the large background of the probe beam.
For instance, when an XFEL probe pulse comprising a large number $N\sim10^{11}$ of photons with an energy of $\omega\sim10\,{\rm keV}$  collides with a tightly focused Petawatt-class laser beam, the attainable quantum vacuum signals are typically below the single photon level per shot; cf. Ref.~\cite{Ahmadiniaz:2024xob} and references therein.
The signal induced in a mode polarised perpendicular to the linearly polarised  probe is the signature of vacuum birefringence.
The requirements for a successful measurement of the nonlinear vacuum response are a twofold: on the one hand the signal itself must be sufficiently large to be detected. On the other hand, the background from the probe beam must be small enough so as not to mask the signal. Ideally the signal should be greater than the background to allow for a significant measurement within a realistic experimental duration.
The experiment must therefore be designed to allow the signal photons to be distinguished from the background. In principle the response of the quantum vacuum can result in signals which differ from the probe beam in terms of their polarisation, angular distribution and photon energy. 
Of particular interest are configurations, that allow the fundamental coupling parameters $a$ and $b$ governing the effective interaction of electromagnetic fields in the underlying theory ($c=\hbar=1$) \cite{Fedotov:2022ely},
\begin{equation}
 {\cal L}_{\rm int}\simeq\frac{m^4}{1440\pi^2}\biggl[a\Bigl(\frac{\vec{B}^2-\vec{E}^2}{E_{\rm S}^2}\Bigr)^2 + b\Bigl(\frac{2\vec{B}\cdot\vec{E}}{E_{\rm S}^2}\Bigr)^2\biggr]\,, \label{eq:Lint}
\end{equation}
to be determined and thus the theoretical framework to be tested in detail ($m$ is the electron mass and $E_{\rm S}=m^2/e=1.3\times10^{18}\,{\rm V}/{\rm m}$ the {\it critical} electric field).
Quantum electrodynamics (QED) predicts $a\simeq4$ and $b\simeq7$ \cite{Euler:1935zz,Heisenberg:1936nmg}.
Higher order corrections are parametrically suppressed by powers of the fine-structure constant $\alpha=1/137$. 
The coupling parameters $a$ and $b$ are also sensitive to physics beyond the Standard Model because \Eqref{eq:Lint} generically emerges as the weak-field limit of theories respecting a charge conjugation parity symmetry.  
Both the parallel and perpendicularly polarised components must be measured to determine $a$ and $b$.
A prospective candidate is the darkfield concept~\cite{Peatross:1994,Zepf:1998,Karbstein:2020gzg} 
utilizing the angular distribution to separate the signal from the background.
Here, we consider the specific implementation put forward in Ref.~\cite{Karbstein:2022uwf} using an XFEL probe colliding with a tightly focused high-intensity pump. 
\onecolumngrid
\begin{center}
\begin{figure}[htp]
    \centering
    \includegraphics[width=\textwidth]{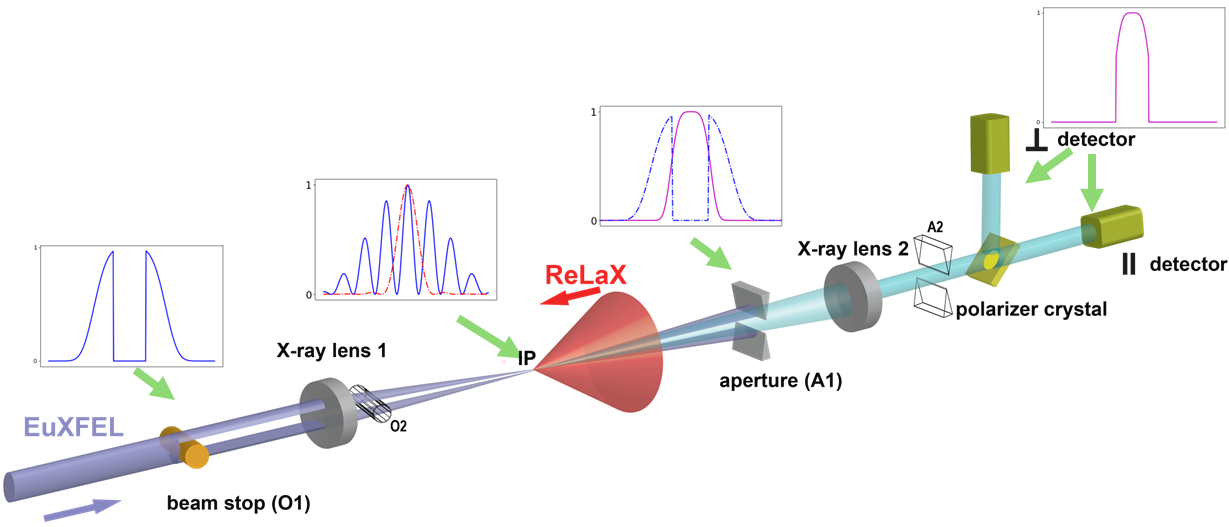}
    \caption{Idealized darkfield set-up with normalized intensity distributions at each critical plane indicated by a green arrow. The primary x-ray distributions (blue) are shown as for 4 planes with the dashed red line showing the laser focal intensity distribution. The central shadow is imprinted onto the x-ray probe beam by inserting an opaque beam stop ('obstacle'), transforming the initial distribution with a central maximum. In the focal plane a peaked distribution reappears with the shadow encoded in side-lobes appearing in the focus. At the interaction-point, IP, the central peak overlaps with the high intensity  laser (ReLaX) focus. For matched x-ray and laser waists the resulting quantum vacuum signal (magenta line in the aperture plane) is peaked on axis, due to the suppression of the side-lobes in the interaction. An additional object O2 immediately after the first lens and matching aperture A2 in the image plane of O2 can also be introduced
    . In the experiment, the detectors are in the focus of lens 2.}
    \label{fig:DF_principle}
\end{figure}
\end{center}
\twocolumngrid
In the following, we detail this measurement concept and report the performance of a recent implementation at the High Energy Density (HED) scientific instrument of the European X-ray Free Electron Laser (EuXFEL) \cite{HED}. Finally, we review the requirements for successful measurement of vacuum birefringence in this set-up.

\section{Darkfield Measurement Concept}

The idea underlying the darkfield (DF) concept is to choose an experimental set-up, where the desired signal is contained in an angular range which is essentially free from any background, thereby maximising the signal to background ratio $\cal R$ \footnote{To our knowledge this is true in the context of two beam interactions. Multi-beam geometries add further possibilities at the cost of added complexity.}. In the specific case of the x-ray DF scenario devised for the measurement of the nonlinear quantum vacuum response, the probe beam is modified with a well-defined obstacle such as to exhibit a shadow in both the converging and expanding beam while retaining a peaked focus profile, see Fig.~\ref{fig:DF_principle} for an illustration.
Wave optics implies that in the focus the information about the shadow in the beam is encoded in pronounced side peaks. 
 The number of x-ray signal photons $N_{\rm sig}$ induced in the collision with the high-intensity pump scales linearly (quadratically) with the intensity of the probe (pump). In line with that, the product of the transverse focus profile of the probe and the squared pump focus profile determines the signal source distribution. Hence, the signal source distribution differs substantially from the probe focus profile and has a different angular distribution; see Fig.~\ref{fig:DF_principle}. 

Appropriate tuning of the pump and probe waists allows the source distribution of the quantum vacuum signal to be modified, in particular the side-lobes can be significantly reduced relative to central peak in the signal source distribution for sufficiently small pump foci. For a probe beam with a central shadow, these side-lobes in its focus encode the shadow imprinted in its near-field.

Effectively erasing the side-lobes in the source distribution of the quantum vacuum signal ensures that the signal does not inherit any information about the shadow in the incident probe beam. Ideally, only the central near-Gaussian peak is left as signal source. This results in an angular signal distribution that is essentially Gaussian and resembles that of the unobstructed probe beam.

In summary, the above approach allows the maximum of the quantum vacuum signal to propagate into the minimum of the shadow in the probe beam, resulting in substantially improved signal to background ratio $\cal {R}$, thus facilitating the detection of the extremely weak quantum vacuum signals.

For sufficiently good background suppression in the shadow and reasonably strong quantum vacuum signals, the DF concept should even facilitate the detection of both the parallel ($\parallel$) and perpendicular ($\perp$) polarised components of the nonlinear vacuum response. This, in turn, would provide direct access to the low-energy constants $a$ and $b$. The key parameter of the DF concept is therefore the quality of the shadow. The latter can be quantified in terms of the unwanted background measured within the shadow. For lossless beam transport we define the shadow factor
associated with a detector of a given detection area $A_{\rm DET}$,
(parametrized by the coordinates $x$ and $y$) as

\begin{equation}
{\cal S}=\frac{1}{N}\int_{A_{\rm DET}}{\rm d}^2x\,\frac{{\rm d}^2N(x,y)}{{\rm d}x\,{\rm d}y}\ .
 \label{eq:shadowquality}
\end{equation}

For a given detector, ${\cal S}$ is the ratio of photons indistinguishable from the quantum vacuum signal to the total number of photons $N$ in the initial x-ray beam.


\section{Experimental Implementation and Design Considerations}

While the outcome of an elementary proof-of-concept experiment of the dark-field approach at an x-ray tube \cite{Karbstein:2022uwf} and the results of numerical diffraction simulations \cite{Ahmadiniaz:2024xob} are promising, reliable information about the shadow factor achievable with an XFEL probe can only be drawn from a full-scale experiment.
To validate the darkfield detection concept, a dedicated x-ray-only proof-of-concept experiment was carried out at the HED scientific instrument of the European XFEL in a beamtime granted by the Helmholtz International Beamline for Extreme Fields (HIBEF) priority access program. This ensures that the beam characteristics are those achievable in a combined experiment and allows a predictive simulation capability to be developed.


In a hypothetical arrangement with perfect imaging and no scattering, the DF would have zero background from the primary XFEL beam in the detector area defined by aperture A1 with the layout in Fig.~\ref{fig:DF_principle}.


In a real-world scenario, every scattering/diffraction event modifies the angular distribution of the x-rays and imaging properties
and the combination of two such events can result in a background x-ray photon propagating along the signal path. Therefore, it is clear that diffracted and scattered x-rays pose the main source of background that must be suppressed. 
Diffraction occurring around the obstacles (O1, O2) results in x-rays propagating at an angle to the unperturbed beam and, as a consequence of that, x-ray photons in the geometric shadow of the blocking obstacles.  
A single further scattering/diffraction event allows such photons to be deflected onto a path leading to the Region of Interest (ROI - typically a single detector pixel) on the detector plane in the focus of lens 2.


Diffraction is suppressed by imaging the obstacles on the matching aperture plane (i.e. O1 images to A1, O2 to A2 and interaction point pinhole to the detector plane). In the limit of perfect imaging, aperture sizes A1, A2 could be chosen to match the size of the obstacles, thus maximizing transmission of the quantum vacuum signal. In practice, image
quality is limited by the small numerical aperture of the
x-ray lenses (lens 1 and lens 2). The point spread function
results in a less sharp  obstacle edge, requiring
smaller apertures to minimise background caused by diffraction from A1, A2. Typically, the apertures were set to a quarter of the size of the image of the obstacle.


Scattering in the Beryllium compound refractive lenses (CRLs) \cite{Lyatun:2020jsr} can also contribute background photons. 


The central area of the first lens, for example, is irradiated by photons diffracted into the geometrical shadow of obstacle O1. Those photons are then scattered among others onto the beam axis, going directly towards detector ROI. To prevent this, the obstacle O2 located just after the lens is introduced. The lens scattering outside of the obstacle shadow, on the other hand, is blocked by the pinhole positioned in the interaction point (IP) between optical and x-ray beams in the focal plane of lens 1. The diameter of the pinhole is chosen so that there is no direct line-of-sight between the brightly illuminated area of lens 1 and opening of A1. Scattering on lens 2 is controlled by making sure that no strong beam is reaching the lens, i.e. the direct part of beam is blocked by A2.

Unless further measures are taken, the scattering in the lens 1 will result in (a) scattering of the diffracted x-rays from O1 into the ROI and (b) homogeneous illumination of lens 2 by scattering from lens 1. In our set-up, scenario (a) is blocked by O2, while scenario (b) is blocked by inserting a pinhole at the interaction point. The combination of O2 and pinhole prevents scattered photons from reaching the lens 2 when A1 is chosen to be smaller than the pinhole shadow. 

In this arrangement a beam photon must be deviated at least {\it three} times through scattering/diffraction to reach the ROI.


\section{Results}\label{sec:results}

To determine the viability of the scheme on the HED Beamline at European XFEL an experiment to characterize the darkfield set-up was performed. This experiment had three main aims. First, to characterize the XFEL properties critical for this kind of experiment. Second, to get scattering data on used components to asses the material properties and guide the simulations to predictive capabilities. And third, to perform the complex setup, demonstrate its feasibility and determine the shadow factor $\cal S$. 
This is particularly important, as scattering and diffraction depend in detail on micro scale surface topology and (non-)uniformity of the bulk composition. Therefore the exact background in the ROI, and subsequently the overall viability of the experiment, can only be determined experimentally.

\subsection{XFEL Beam Properties}

The XFEL was operated in the hard x-ray self seeded mode (HXRSS) \cite{Liu:2019cpe}. 
Compared to the SASE regime, this results in a reduction in total photon number while increasing the flux in a narrow (sub-eV) bandwidth.  Narrow bandwidth is essential to achieve well defined focal quality with minimal chromatic aberrations and to maximise the signal within the angular acceptance (and therefore reflection bandwidth) of the crystal polarizing optics used in the detection set-up.

\begin{figure}[htp]
    \centering
    \includegraphics[width=0.99\linewidth]{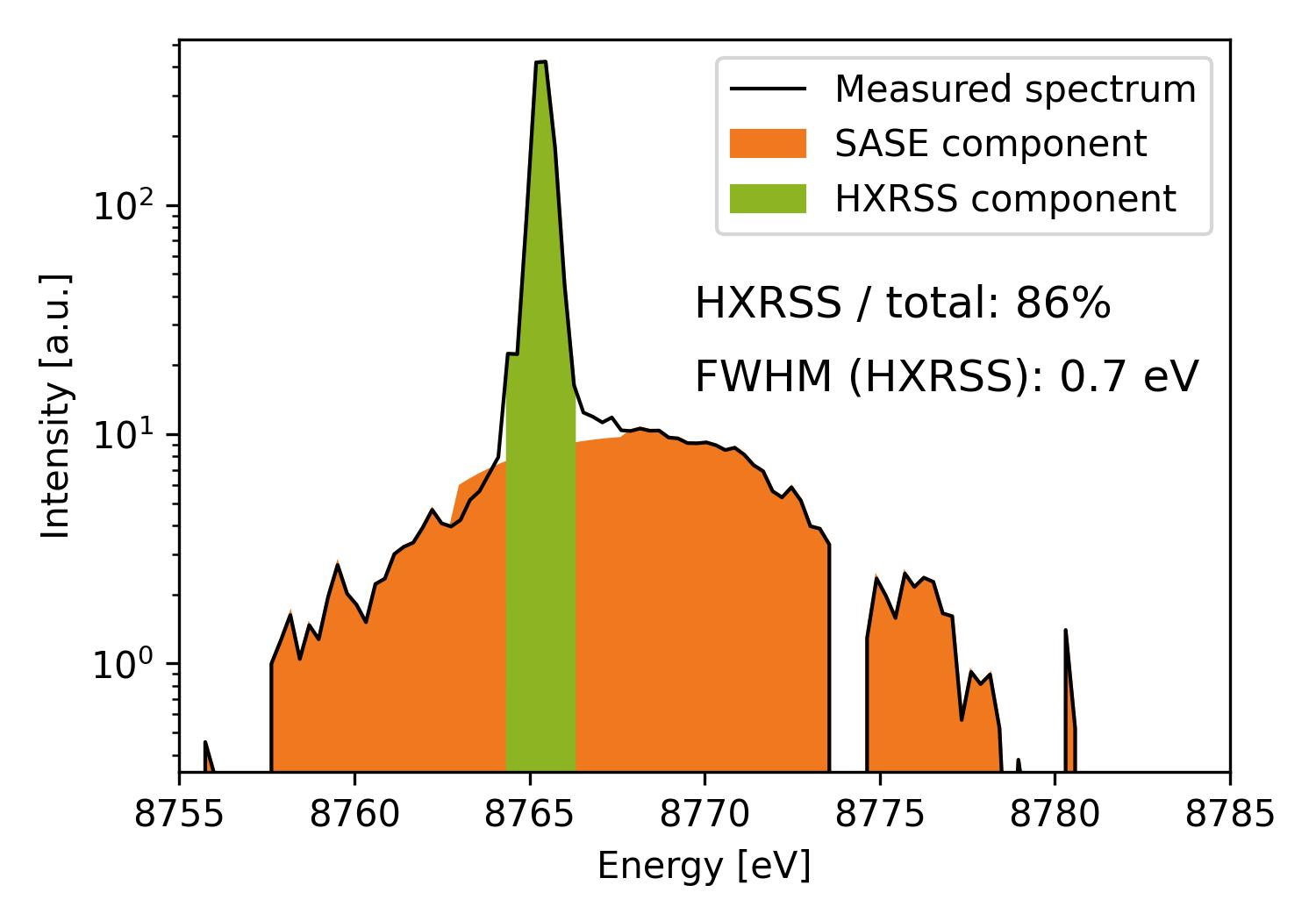}
    
    \caption{XFEL spectrum measured by the Hirex spectrometer. The SASE bandwidth is \SI{25}{eV} with seeded peak positioned at \SI{8766}{eV} with  FWHM \SI{0.7}{eV} containing $86\%$ of the beam energy.
}
    \label{fig:hirex_spect}
\end{figure}

The EuXFEL delivered beam had pulse energies in the range of $W_{\rm x} = 300-500\,\upmu$J in seeded mode, as measured by gas monitors before entering the HED instrument beamline. Fig. \ref{fig:hirex_spect} shows the XFEL spectrum measured with the HIREX spectrometer \cite{Kujala:2020rsi}.
The seeded peak (FWHM $<\SI{1}{\eV}$) set at  $\omega = 8766\,{\rm eV}$ ($\lambda=0.141\,{\rm nm}$) is clearly visible above the SASE background ($\approx25\,{\rm eV}$ bandwidth). It contains 86\% of the total pulse energy.

\begin{figure}[htp]
    \centering
    \includegraphics[width=0.85\linewidth]{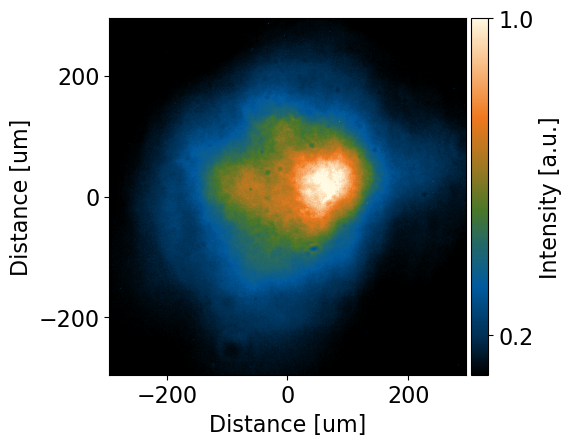}
\caption{Collimated XFEL beam profile on the detector plane.}
    
    \label{fig:NF_Data}
\end{figure}

\begin{figure}[htp]
    \centering
    \includegraphics[width=0.99\linewidth]{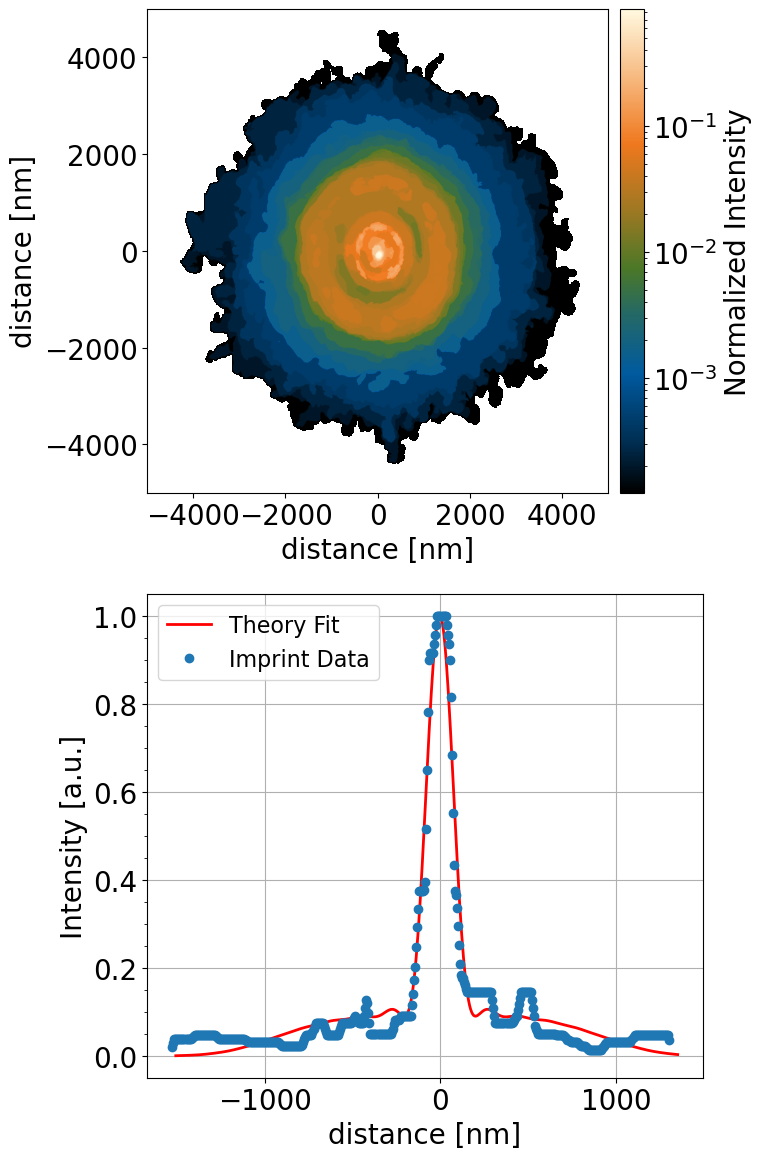}
    \caption{Focal intensity distribution of the full XFEL beam measured with ablation imprints (top panel). The bottom panels shows a  lineout through the centre of the imprint (blue dots), and a theoretical fit (red line). The fit is an Airy focus with central peak diameter ($2w_0$) of \SI{240}{\nm} to which a small super Gaussian background representing the scattering background in the focusing lenses is added; see main text.}
    \label{fig:focus_full}
\end{figure}

The XFEL beam shown in Fig.~\ref{fig:NF_Data} 
was focused using a Beryllium CRL with a focal length of $f=438$\,mm 
measured from the middle of the 13-element lens stack. Predicting the quantum vacuum signal in this set-up requires detailed knowledge of the XFEL focus. This was characterized using specially designed and characterized ablation targets, consisting of highly damage resistant diamond substrate with a well-defined \ch{PbI2} coating serving as the ablation material \cite{Imprint:2013, Imprint:2022}. Fig.~\ref{fig:focus_full} shows the composite image of multiple ablation shots in the focal position. 
The analytic approximation of the focus intensity profile ('Airy focus') $I(r) = I_0\bigl[J_1\bigl(2\sqrt{2(1-e^{-1})}\,r/w_0\bigr)/\bigl(\sqrt{2(1-e^{-1})}\,r/w_0\bigr)\bigr]^2$ of a rotationally symmetric beam with a flat top transverse profile in its near-field was fitted to the data. Here, $I_0$ is the peak intensity, $J_n$ is the Bessel function of the first kind of order $n$, and $w_0$ is the beam waist at approximate $1/e^2$ \cite{Karbstein:2023}.


The broad-band SASE background visible in Fig.~\ref{fig:hirex_spect} is focused at different positions along the focal axis due to lens chromaticity and adds a low intensity pedestal at the focus of the seeded peak. The latter is fitted by a super Gaussian function $\sim\exp\{-1/2\,(r/\sigma)^4\}$ with $\sigma=\SI{850}{\nano\meter}$ and peak intensity of about $I_0/10$ and is added to the Airy focus fit in Figs ~\ref{fig:focus_full} and ~\ref{fig:focus_obs}. The fit  yields a width of the central peak ($2w_0$) of \SI{240}{\nm} (140 nm FWHM), demonstrating  near diffraction limited focusing. The diffraction limited value is $2w_0\simeq2\lambda(f/d)\sqrt{8(1-e^{-1})}/\pi=227\,{\rm nm}$ \cite{Karbstein:2023}. The data shown in the following was recorded with an average XFEL beam energy of \SI{108}{\micro\joule} measured after lens 1, we obtain an effective photon number of $N_0=\SI{6.58e10}{}$ in the central Airy peak containing $68.8\%$ at $I>I_0/e^2$ in this run. The photon number $N_0$ in the collimated beam without obstacle but corrected for lens losses  is the input into our theoretical calculations of the expected quantum vacuum signal in Section~\ref{sec:discussion} below.

\subsection{Focal Distribution with Beam-Stop Object}

As described above, the key steps to implementing the darkfield measurement set-up are first to imprint a high-quality shadow in the EuXFEL beam and then to interact a tightly focused high intensity laser with the x-ray far-field intensity pattern the lens focus, resulting in quantum vacuum signal with strong signal in the shadow.
The central shadow was created using a polished tungsten wire with a diameter of \SI{160} \um\, as obstacle (Fig.~\ref{fig:DF_principle}), obscuring about half of the beam ($52.2\%$ of the beam area). 

\begin{figure}[htp]
    \centering
    
   \includegraphics[width=0.99\linewidth]{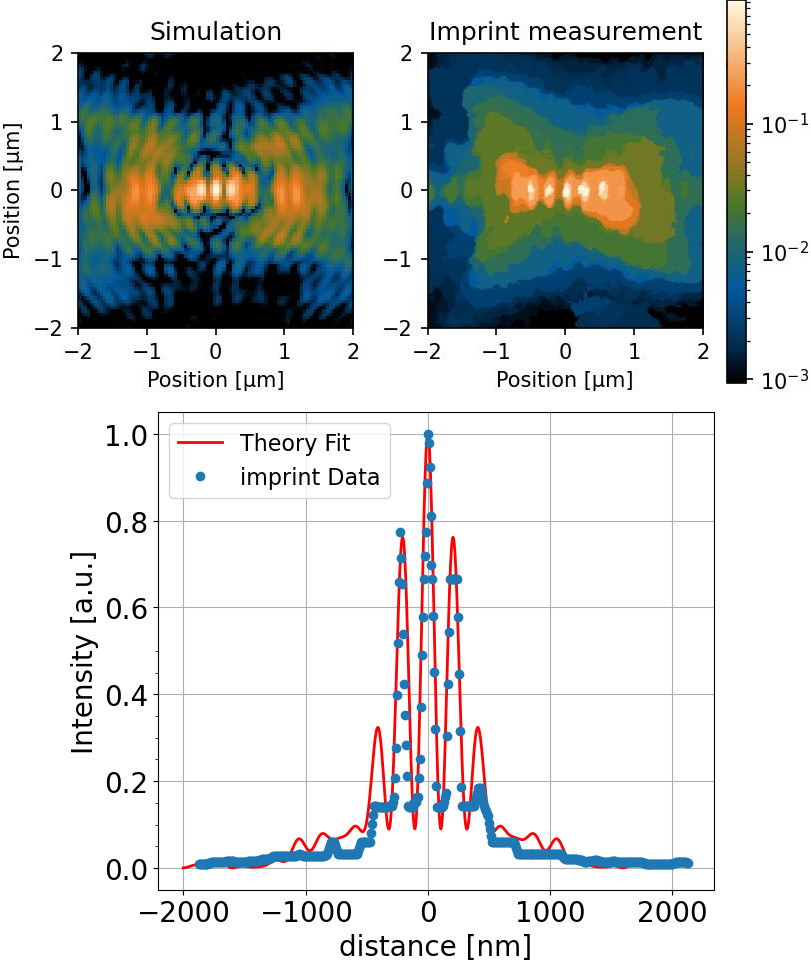}

    \caption{
    Focal distribution of XFEL beam with \SI{160}{\micro\meter}  electro-chemically polished tungsten wire as obstacle O1 and another as O2 after lens 1, simulation (top left panel), and reconstruction from the imprint measurement (top right). Both normalized. (lower panel) the central line out of the imprint measurements (blue dots), and the theoretical fit (red line) yielding an $1/e^2$ width ($2w_0$) of \SI{180}{\nano\meter} for the central peak. 
    }
    \label{fig:focus_obs}
\end{figure}

The shadow of the wire results in fringes in the focus of the beam. The experimentally observed pattern with a periodicity of $0.2\,\upmu{\rm m}$  is shown in Fig.~\ref{fig:focus_obs}.

The analytic approximation of the focal intensity profile of a beam with flat top near-field profile (width $d$) featuring a central rectangular shadow (width $d_{\rm wire}$) is 
$I(x)
=I_0\cos^2\bigl(\frac{(1+\nu)f(\nu)}{2}\frac{x}{w_0}\bigr)\,{\rm sinc}^2\bigl(\frac{(1-\nu)f(\nu)}{2}\frac{x}{w_0}\bigr)$, where $x$ is the coordinate perpendicular to the wire, $\nu = d_{\rm wire}/d$, and $f(\nu) = \sqrt{6(1-e^{-1})(1-\nu)/(1-\nu^3)}$ \cite{Karbstein:2025}
This profile is fitted to the data and yielded a central peak diameter ($2w_0$) of \SI{180}{\nm} fitted at approximate $1/e^2$ compared to the data. 

The same super Gaussian function as for the case of the unobstructed XFEL beam is added to model the low intensity SASE background.
Small discrepancies between measured data and theory in the higher order (side) peaks are visible. 

\subsection{Determination of Shadow Factor}

The shadow factor ${\cal S}$ used in the theoretical considerations can be derived from the experimental ratio between the signal in the ROI containing the image of the focal spot with and without blocking object(s) in place,
which we denote ${\cal S}_{\rm exp}$. The detector is placed in the image of the XFEL focus created by lens 2.\\
The experimental ratio ${\cal S}_{\rm exp}$ is
 related to the shadow factor ${\cal S}$ in \Eqref{eq:shadowquality} as ${\cal S}= T_{A1} \times {\cal S}_{\rm exp}$, with $T_{A1}$ 
the transmission at the aperture $A_1$.

Figure  6 shows an image taken with a high resolution scintillator camera at the image plane. By design, the most intense
diffracted/scattered light is visible outside the image of
the focus with the image of the focus restricted to one
pixel in the centre of the distribution.
From the distribution of the signal in Fig. 6, it is clear
that the background signal depends on pixel size of the
detector (Jungfrau: 75 µm, Andor IKON 13 µm). The image of the focus is typically smaller than the pixel size).
Most scans were performed with the Jungfrau
detector\cite{Jungfrau:2023}, due to its single photon sensitivity, large dynamic range and fast acquisition. Figure 7 shows ${\cal S }$ as
a function of the aperture sizes $A_1$, $A_2$ for fixed obstacle
size with each data point being the average of typically
110 shots. Values of ${\cal S}_{\rm exp} \sim 10^{-9}$ were observed over
a range of aperture settings. To determine the improvement of ${\cal S}_{\rm exp}$ with the smaller-pixel  detector
envisaged for the full experiment (nominally reducing the
ROI area to 3$\%$), the signal distribution was recorded using
the Andor iKON camera. The measurement (Fig. 8) gives a slightly better reduction in background to 1.97$\%$, as a result of the inhomogeneous background distribution shown in Figures  6.

Correcting the  ${\cal S}_{\rm exp}$ measurement for pixel size and $T_{A1}$ we derive shadow factors of ${\cal S} < 3 \times 10^{-11}$ from our data for cameras with small pixels for practicable aperture settings.

\begin{figure}[htp]
    \centering
    \includegraphics[width=0.95\linewidth]{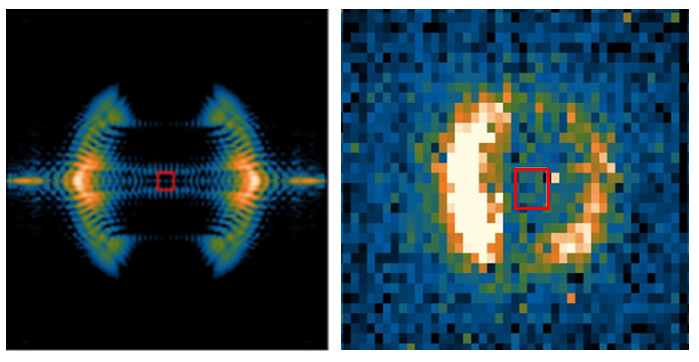}
    \caption{Comparison of image on detector plane from simulations (left) and experiment  with high resolution scintillator (right). The bright feature is scattering from the \SI{25}{\micro\meter} tungsten pinhole illuminated by scattered and diffracted x-rays. The red square is 13 \um\, ROI.}
    \label{OptiquePeter}
\end{figure}

\begin{figure}[htp]
    \centering
    \includegraphics[width=1\linewidth]{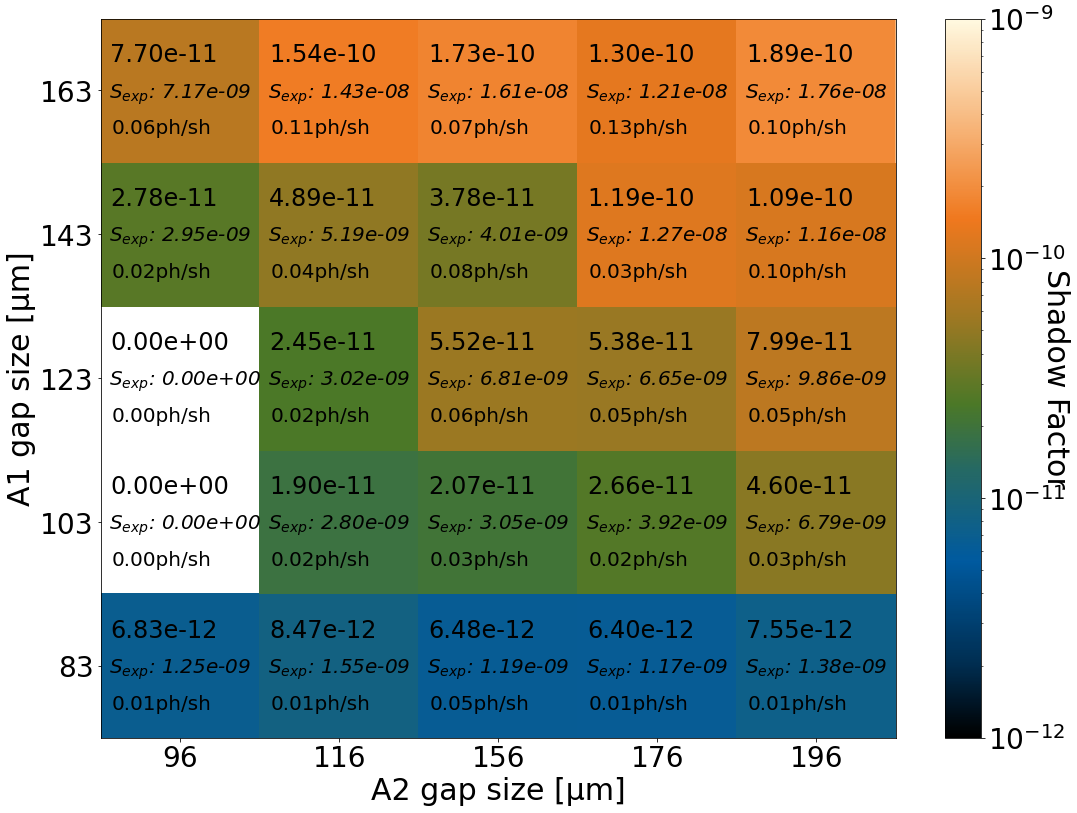}
    \caption{Results as a function of the aperture sizes A1 and A2. Numbers in each box are the shadow factor $\cal S$ (top number) for a pixel size $13\,\upmu{\rm m}$ which is estimated from the experimentally derived ratio ${\cal S}_{\rm exp}$ measured using the $75\,\upmu{\rm m}$ pixel Jungfrau detector (middle). Average photon number per shot (bottom) registered in a single $75\,\upmu{\rm m}$ pixel. 
    }
    \label{fig:JF_SF}
\end{figure}



Experimentally derived shadow factors calculated for  the planned 13\ \um\ pixel size are shown in Fig. \ref{fig:JF_SF} as a function of the aperture width A1, A2 for configuration with obstacle diameters 180 and $160\,\upmu{\rm m}$ for O1 and O2, respectively. Each each data point is the average of typically 110 shots. The values measured using 75 \um\ pixels are labelled as  ${\cal S}_{\rm exp}$ in the figure. 


\begin{figure}
        \centering
        \includegraphics[width=0.95\linewidth]{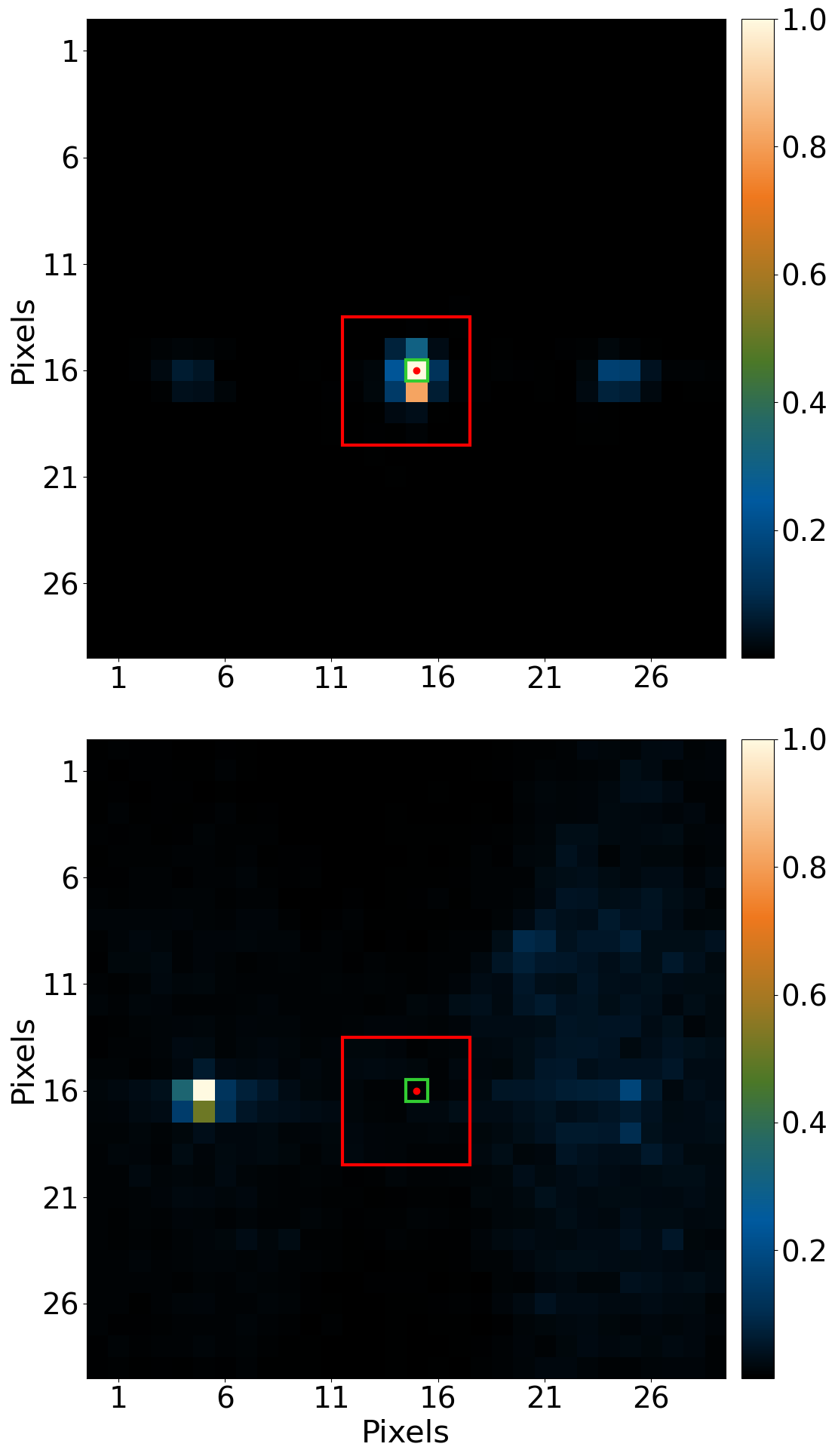}
    \caption{
    Focused XFEL beam (above) and darkfield measurement (below) on iKon CCD. The red dot marks our ROI (framed by the lime square) in both images. The red square is the Jungfrau detector pixel size. 
    The ROI on iKon CCD with \num{13}$\times$\SI{13}{\micro\metre\squared} contains only \SI{1.97}{\percent} background compared to the ROI on Jungfrau with \num{75}$\times$\SI{75}{\micro\metre\squared}.}
        \label{fig:IKONimages}
\end{figure}

\subsection{Comparison to Simulations}

\begin{figure}
        \centering
        \includegraphics[width=1\linewidth]{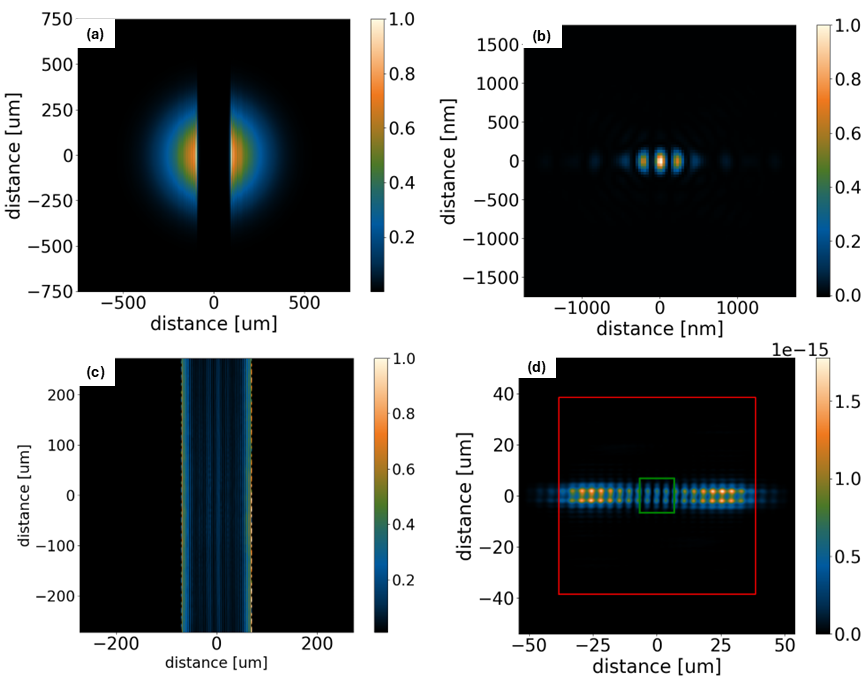}
    \caption{Simulated transverse profiles of the beam obstructed with beam blocks O1 (diameter \SI{180}{\micro\metre}) and O2 (diameter \SI{160}{\micro\metre}) in 4 critical plane. (a) Before the first lens, (b) in the focus, (c) after aperture A1 with \SI{140}{\micro\metre} gap, and (d) on the detector. The red square represent the ROI on Jungfrau with \num{75}$\times$\SI{75}{\micro\metre\squared}.}
        \label{fig:simulation}
\end{figure}

To confirm our understanding of the darkfield setup,  the experimental geometry was simulated in a diffraction code based on the \textit{LightPipes} package \cite{LightPipes:1997,LightPipes:2.1.5}. Figure~\ref{fig:simulation} shows the simulated transverse beam profiles in four critical planes. These simulations use the experimental parameters used for the data shown in Fig.~\ref{fig:JF_SF}: Tungsten wires as obstacles O1 (diameter \SI{180}{\micro\metre}) and O2 (diameter \SI{160}{\micro\metre}), and apertures A1 (gap size \SI{140}{\micro\metre}) and A2 (gap size \SI{100}{\micro\metre}). 

The simulations account for diffraction effects and include known phase defects of the lenses \cite{Seiboth:me6131,Celestre}. 

The  ${\cal S}_{\rm exp}$ shadow factor in the simulations, Fig.~\ref{fig:simulation} (d), ${\cal S}_{\rm exp}=\num{2.83e-9}$   matches  the experimental value, \num{2.95e-9} well (see Fig.~\ref{fig:JF_SF}). 
Overall it is found that the simulation results agree very well for the bright features. The object of the simulations here is challenging, in that we are accurately trying to predict areas of the weakest signal. Unsurprisingly, this depends sensitively on simulation details, both numerical and the precise set-up of the objects. For example, assuming perfect edges for aperture A1 overestimated the signal level.  Adding (estimated) imperfections to A1 aperture edges improved the agreement with the experiment. Nonetheless, the current code has been shown to have sufficiently good agreement to be used to guide set-up development.

\subsection{Polarisation Selection}

To determine the coupling parameters $a$ and $b$, we use a polarisation-selective beam splitter using the Ge 440 crystal plane. The crystal was setup so that each polarization component is reflected on different of two orthogonal crystal planes. This ensures that each polarization component is reflected $90^\circ$ to each other. 
This principle was first explained by Baranova and Stepanenko for hexagonal crystals \cite{Baranova} and further developed for cubic crystals by Wallace, Presura and Haque  \cite{Wallace2020, Wallace2021, Presura}.  The crystals were asymmetrically, such that  the incoming beam is incident on the crystal surface closer to grazing incidence, but still at $45^\circ$ to the crystal planes, ensuring increased reflection bandwidth and therefore higher integrated reflectivity.

We define polarisation purity as the ratio of signal after the polarisation analyser set to measure perpendicular polarisation to the total beam intensity.  This measured value can be limited either by the polarisation of the incoming beam, or by the selective power of the crystal.

\begin{figure}
        \centering
        \includegraphics[scale=0.6]{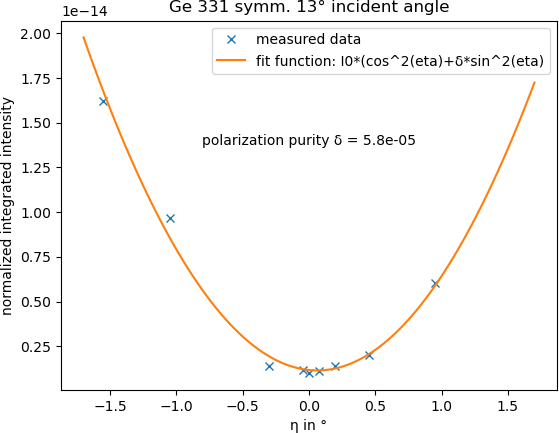}
    \caption{Extinction curve with \ch{Ge}-331 analyser. polarisation purity measurement with a symmetric \ch{Ge}-331. The analyser set-up has a transmission of 21.4\%. }
        \label{fig:pol_willi}
\end{figure}

The measurement with a crystal cut so that a beam was incident the crystal at $29^\circ$ to its surface yielded polarisation purity ${\cal P}=6.8\times10^{-5}$ and transmission $T=5\%$.
Increasing the asymmetry to incidence angle $13^\circ$ kept the purity at similar level, ${\cal P}=5.8\times10^{-5}$, but the transmission increased to $T=12\%$, see 
Fig.~\ref{fig:pol_willi}. 

Further polarisation purity measurements has shown the polarisation degradation due to the \ch{Be}-lenses are negligible. Using such an analyser can thus effectively improve the shadow quality ${\cal S}$ for measurements of the $\perp$-polarised component of the nonlinear quantum vacuum response down to ${\cal SP}<10^{-15}$.

The polarisation purity ${\cal P}$ was determined by rotating the analyser around the x-ray beam ($\eta$-circle) close to the extinction and measuring a rocking curve at each position. The dependence of the integrated intensity of the rocking curves on $\eta$ is given by Malus's law
\begin{equation}
    I(\eta) \propto I_0\bigl(\sin^2\eta+{\cal P}\cos^2\eta\bigr) \,.
\end{equation}
The transmission of the analyser can be measured with the $\perp$ or $\parallel$ polarised component of the x-ray beam. During the beam time it was figured out that the detector used to measure the $\parallel$ polarised component exhibited a huge spatial inhomogeneity. Therefore, we determine the transmission of the analyser crystal with the $\perp$ polarised component as
\begin{equation}
    T_{\rm analyser} = \frac{I(\eta=0^\circ)}{I_0\times T_{\rm absorbers} \times T_{\rm air} \times {\cal P}} \,,
\end{equation}
where $I(\eta=0^\circ)$ represents the measured intensity at the detector in the extinction position, $I_0$ is the intensity in front of the analyser, $T_{\rm absorbers}$ is the transmission of the aluminium absorber foils right in front of the detector and $T_{\rm air}$ is the transmission of the remaining $1662\,{\rm mm}$ air in the beam path.

\section{Discussion} \label{sec:discussion}

\begin{figure}[htp]
    \centering
    \includegraphics[width=0.9\linewidth]{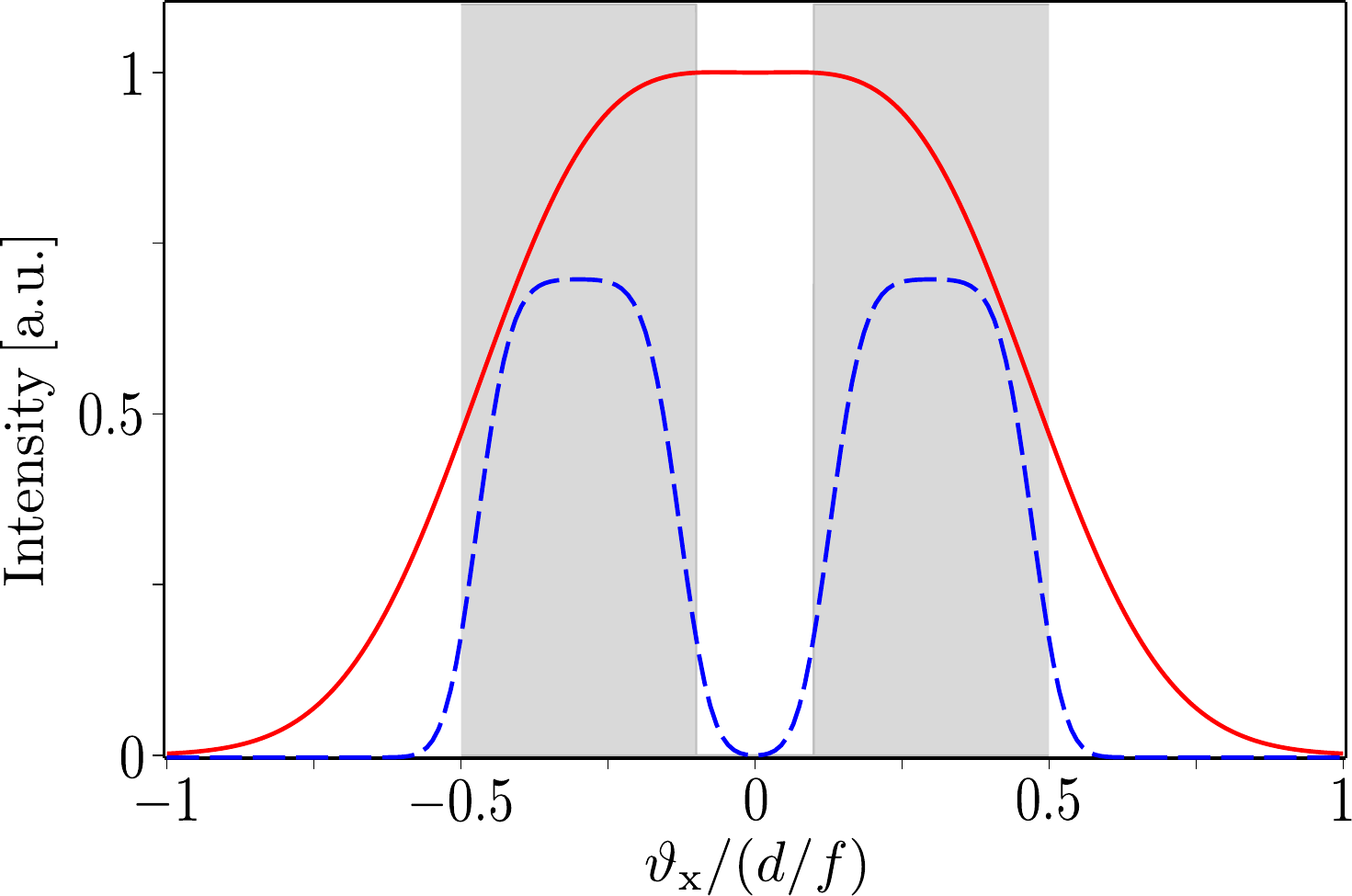}
    \caption{Distribution of the predicted signal far away from the focus in the direction perpendicular to the obstacle; the blocking fraction is $d_{\rm wire}/d=20\%$, and the angle $\vartheta_{\rm x}$ is measured in units of $d/f$. The gray-shaded areas mark the directions of the bright probe beam and the central white area corresponds to the 'shadow' of the obstacle.
    The solid red curve for $d/f=140\,\upmu{\rm rad}$ is matched to the focal spot of the high intensity laser with $w_{\rm laser}\approx 2 w_0$; cf. Fig.~\ref{fig:Felix_Plot}. The dashed blue curve for $d/f=700\,\upmu{\rm rad}$ illustrates that the signal inherits the 'shadow' feature from the probe beam for $w_{\rm laser}\gg w_0$.}
    \label{fig:TheoryAngDist}
\end{figure}

The above results focus on optimizing the shadow factor. Clearly, the trivial optimum of ${\cal S}$ would occur for closed apertures A1, A2. The main constraint on a real-world set-up comes from matching the x-ray spatial scale of the x-ray intensity distribution to the size of the high-intensity laser focus. 
In the limit where the waist of the optical laser $w_{\rm laser} \gg w_0$, the distribution of the quantum vacuum signal in focus is identical to that of the XFEL beam and therefore the signal on axis is not increased resulting in the quantum vacuum signal being suppressed by essentially the same factor $\cal S$ as the background. For the matched case we find $w_{\rm laser} \approx 2 w_{\rm 0}$ resulting in only one dominant peak in the signal distribution in the source point and therefore a peaked distribution of the quantum vacuum signal at the position of A1. These two cases are illustrated in Fig.~\ref{fig:TheoryAngDist}.
The combined problem of suppressing the background due to diffraction/scattering while optimizing the quantum vacuum signal is considered for an x-ray photon energy of $\omega=8766\,{\rm eV}$ (FWHM pulse duration $25\,{\rm fs}$), optical beam waist $w_{\rm FWHM}=1.3\,\upmu{\rm m}$ (wavelength $800\,{\rm nm}$, FWHM pulse duration $30\,{\rm fs}$) expected to be attainable with $f/1$ focussing \cite{Ahmadiniaz:2024xob,ReLaX}, a detector acceptance angle due to aperture A1 of $\Theta_{\rm det}=\Theta_{\rm wire}/4$. For very small or very large wire diameters the signal trivially goes towards zero. The matching between the $w_{\rm laser}$ and $ w_0$ is calculated by varying the XFEL focusing parameter $d/f$ where, $d$ is the lens beam size and $f$ the CRL focal length. It is seen that the optimal values lies in the range $d/f = 100\dots200 ~ \upmu \mathrm{rad}$.
For larger values, 
the pump focus is much larger then the probe beam focus, therefore the quantum vacuum signal obtains a spatial distribution near-identical to that of the probe beam with a central minimum (shadow) and gets blocked on aperture A1. 
\begin{figure}[htp]
    \includegraphics[width=1\linewidth]{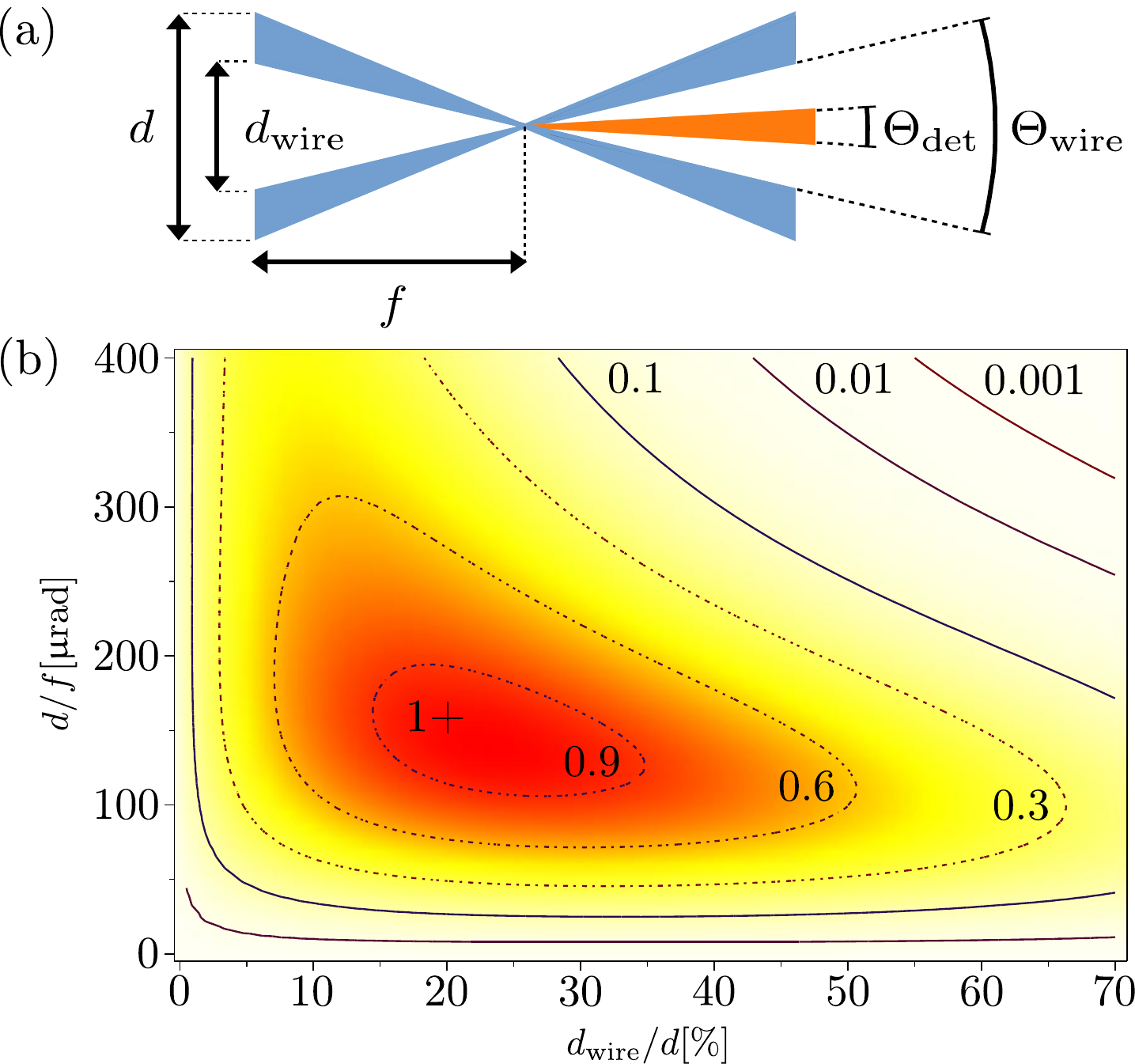}
    \caption{(a) Sketch of the setting and relevant quantities. (b) Normalized quantum vacuum signal as a function of the blocking fraction $d_{\rm wire}/d$ and the divergence $d/f$ of the x-ray beam; x-ray photon energy $\omega=8766\,{\rm eV}$, optical beam waist $w_{\rm FWHM}=1.3\,\upmu{\rm m}$ and detector acceptance angle $\Theta_{\rm det}=\Theta_{\rm wire}/4$. The factors labelling the isocontours indicate the drop of the signal from its maximum value marked by a cross at $d_{\rm wire}/d=20\%$ and $d/f=160\,\upmu{\rm rad}$.}  
    \label{fig:Felix_Plot}
\end{figure}

For the optimal parameters identified in Fig.~\ref{fig:Felix_Plot}, we obtain 
\begin{equation}
    N_{\parallel,\perp}\simeq 2.2\times10^{-17}\,c_{\parallel,\perp}\Bigl(\frac{W}{1\,{\rm J}}\Bigr)^2 N
    \label{eq:signals}
\end{equation}
signal photons. Here, $W$ is the pulse energy of the optical laser and $N$ is related to the number $N_{\rm probe}$ of focused XFEL photons available for probing the vacuum polarised by the pump pulse as $N=N_{\rm probe}/(1-d_{\rm wire}/d)$. The coefficients $c_{\parallel,\perp}$ depend on $a$, $b$ and the relative polarisation $\phi$ of pump and probe.
For $\phi=\pi/4$ maximising the polarisation-flip signal these simplify to $c_\parallel=(a+b)^2\simeq121$ and $c_\perp=(a-b)^2\simeq9$, where the numerical values are the leading-order QED predictions; cf. also \cite{Karbstein:2022uwf}.
The experimental confirmation of a signal $N_{\rm sig}$ in the presence of a background $N_{\rm bgr}$  depends sensitively on the ratio ${\cal R}=N_{\rm sig}/N_{\rm bgr}$. Achieving a significance of $\#\sigma$ requires the number of successful shots $n$ to fulfil \cite{Cowan:2010js}
\begin{equation}
    nN_{\rm sig} > \#^2\,\frac{1}{2}\bigl[\bigl(1+{\cal R}^{-1}\bigr)\ln\bigl(1+{\cal R}\bigl)-1\bigr]^{-1}\,.
    \label{eq:nshot}
\end{equation}

The best shadow factor demonstrated above is
${\cal S}=6.8\times10^{-12}$. This implies that the background for the $\parallel$ mode consists of $N_{\rm bgr}={\cal S}N$ photons. Adopting this choice for the $\parallel$ signal in \Eqref{eq:signals} with $W=4.8\,{\rm J}$, we obtain ${\cal R}=9.0\times10^{-3}$. Hence, in this case \Eqref{eq:nshot} can be translated into the requirement $n > \#^2\,1.5\times10^{15}/N_{\rm probe}$.

Accounting for the additional polarisation suppression by a factor of ${\cal P}=5.8\times10^{-5}$, for the $\perp$ mode we have $N_{\rm bgr}={\cal S}{\cal P}N$. For the $\perp$ signal in \Eqref{eq:signals} with $W=4.8\,{\rm J}$ we thus find ${\cal R}=11.6$, and arrive at the criterion $n > \#^2\,5.0\times10^{13}/N_{\rm probe}$.
Also note that in order to reduce the threshold on $nN_{\rm probe}$ for the $\parallel$ signal to the one obtained for the $\perp$ signal, assuming fixed other parameters, the pump energy has to be increased to $W=11.2\,{\rm J}\approx2.3\times 4.8\,{\rm J}$.

The number of photons $N_0$ contained in the central Airy peak of $1/e^2$ width is related to the total number of photons $N$ in the focus of the unobstructed beam as $N_0/N=1-J_0^2(2\sqrt{2(1-e^{-1})})-J_1^2(2\sqrt{2(1-e^{-1})})=69\%$.
Hence, the value of $N_0$ in the present proof-of-principle experiment extracted in Sec.~\ref{sec:results} implies $N_{\rm probe}=7.57\times10^{10}$ for an optimal blocking of $d_{\rm wire}/d=20\%$ and current EuXFEL performance. 
With this value the above requirements on the numbers of successful shots become $n>\#^2\,2.0\times10^5$ for the $\parallel$ signal and $n>\#^2\,6.6\times10^2$ for the $\perp$ signal.
We emphasize that these conditions could be substantially improved by increasing the pulse energy of the optical laser. For instance, increasing the laser pulse energy by a factor of ten would decrease the threshold on $n$ for the $\parallel$ signal by a factor of $\approx7.9\times10^3$ and that for the $\perp$ signal by a factor of $\approx3.5\times10^2$.

\section{Conclusion and Outlook}

The dark-field concept is a highly promising variant on the road to direct experimental studies of the of the optical response of the quantum vacuum. The concept promises a  detectable signal level despite the large number of probe photons that can contribute to unwanted background and the small cross-sections.  The dark-field approach allows the determination of the fundamental constants governing non-linear quantum vacuum interactions for measurements of both parallel and perpendicularly polarised signal components and, at reduced experimental demands, a measurement of the perpendicularly polarised component only.

\section{Data Availability}

Data recorded for the experiment at the European XFEL are available at \cite{raw-data-5438}\cite{raw-data-6436}.

\begin{acknowledgments}
We acknowledge the European XFEL in Schenefeld, Germany, for provision of X-ray free-electron laser beamtime under the HIBEF priority access route at the Scientific Instrument HED (High Energy Density Science) under proposal numbers 5438 and 6436. We are grateful for assistance and dedication of the XFEL and HIBEF staff that enabled this experiment.
This work has been funded by the Deutsche Forschungsgemeinschaft (DFG, German Research Foundation) under Grants Nos.\ 392856280, 416607684, 416702141, and 416708866 within the Research Unit FOR2783/2.
\end{acknowledgments}



\begin{thebibliography}{10}

\bibitem{Ahmadiniaz:2024xob}
N.~Ahmadiniaz, \textit{et al.},
High Power Laser Sci. Eng. \textbf{13}, e7 (2025)
[arXiv:2405.18063 [physics.ins-det]].

\bibitem{Fedotov:2022ely}
A.~Fedotov, A.~Ilderton, F.~Karbstein, B.~King, D.~Seipt, H.~Taya and G.~Torgrimsson,
Phys. Rept. \textbf{1010}, 1-138 (2023).

\bibitem{Euler:1935zz}
H.~Euler and B.~Kockel,
Naturwiss. \textbf{23}, no.15, 246-247 (1935).

\bibitem{Heisenberg:1936nmg}
W.~Heisenberg and H.~Euler,
Z. Phys. \textbf{98}, no.11-12, 714-732 (1936).

\bibitem{Peatross:1994}
J.~Peatross, J.~L.~Chaloupka and D.~D.~Meyerhofer,
Optical Letters \textbf{19}, 942-944 (1994).

\bibitem{Zepf:1998}
M.~Zepf, \textit{et al.},
Phys. Rev. E \textbf{58}, R5253-R5256 (1998).

\bibitem{Karbstein:2020gzg}
F.~Karbstein and E.~A.~Mosman,
Phys. Rev. D \textbf{101}, 113002 (2020).

\bibitem{Karbstein:2022uwf}
F.~Karbstein, D.~Ullmann, E.~A.~Mosman and M.~Zepf,
Phys. Rev. Lett. \textbf{129}, 061802 (2022).

\bibitem{HED}
U.~Zastrau, \textit{et al.},
J. Synchrotron Radiat. {\bf 28}, 5, 1393-1416 (2021).

\bibitem{Lyatun:2020jsr}
I.~Lyatun, \textit{et al.},
J. Synchrotron Radiat. {\bf 27}, no.1, 44–50 (2020).

\bibitem{Liu:2019cpe}
S.~Liu, W.~Decking, V.~Kocharyan, E.~Saldin, S.~Serkez, R.~Shayduk, H.~Sinn and G.~Geloni,
Phys. Rev. Accel. Beams \textbf{22}, 060704 (2019).

\bibitem{Kujala:2020rsi}
N.~Kujala, \textit{et al.},
Rev. Sci. Instrum. \textbf{91}, 103101 (2020).

\bibitem{Imprint:2013}
J.~Chalupsk\'y, \textit{et al.},
Opt. Express {\bf 21}, no.22, 26363-26375 (2013).

\bibitem{Imprint:2022}
G. ~Mercurio, 
\textit{et al.}, Opt. Express {\bf 30}, 20980-20998 (2022).

\bibitem{Karbstein:2023}
F.~Karbstein,
Habilitation thesis, Faculty of Physics and Astronomy, Friedrich-Schiller-Universit\"at Jena (2024)
\url{https://doi.org/10.22032/dbt.59618}.

\bibitem{Karbstein:2025}
F.~Karbstein, E.~A.~Mosman and M.~Zepf,
{\it to be published} (2025).

\bibitem{Jungfrau:2023}
M.~Sikorski, \textit{et al.},
Frontiers in Physics \textbf{11}, 1303247 (2023).

\bibitem{Gawne:2024}
T.~Gawne, \textit{et al.},
Phys. Rev. B. \textbf{109}, L241112 (2024).

\bibitem{LightPipes:1997}
G.~Vdovin, H.~van~Brug and F.~van Goor,
Proc. SPIE {\bf 3190}, 82 - 93 (1997).

\bibitem{LightPipes:2.1.5}
G.~Vdovin and F.~van Goor,
\url{https://opticspy.github.io/lightpipes/}.

\bibitem{Baranova}
E.~O.~Baranova, M.~M.~Stepanenko,
Plasma Phys. Control. Fusion \textbf{45}, 1113 (2003).

\bibitem{Wallace2020}
M.~S.~Wallace, 
\textit{et al.},
Rev. Sci. Instrum. \textbf{91}, 023105 (2020).

\bibitem{Wallace2021}
M.~S.~Wallace, 
\textit{et al.},
Rev. Sci. Instrum. \textbf{92}, 103101 (2021).

\bibitem{Presura}
R.~Presura, 
\textit{et al.},
Rev. Sci. Instrum. \textbf{92}, 073102 (2021).

\bibitem{ReLaX}
A.~Laso~Garcia, \textit{et al.},
High Power Laser Sci. Eng. {\bf 9}, e59 (2021).

\bibitem{Cowan:2010js}
G.~Cowan, K.~Cranmer, E.~Gross and O.~Vitells,
Eur. Phys. J. C \textbf{71}, 1554 (2011)
[erratum: Eur. Phys. J. C \textbf{73}, 2501 (2013)].

\bibitem{Seiboth:me6131}
F.~Seiboth, \textit{et al.},
J. Synchrotron Rad. \textbf{25}, 108{--}115 (2018).

\bibitem{Celestre}
R.~Celestre, \textit{et al.},
J. Synchrotron Rad. \textbf{27}, 305 (2020).


\bibitem{raw-data-5438}
Data publication, HIBEF PA - Towards Quantum Vacuum Birefringence using an X-ray Dark-field Technique,
\url{https://doi.org/10.22003/XFEL.EU-DATA-005438-00}.

\bibitem{raw-data-6436}
Data publication, HIBEF PA - Next Step towards Vacuum Birefringence using an X-ray Dark-Field Technique,
\url{https://doi.org/10.22003/XFEL.EU-DATA-006436-00}.


\end{thebibliography}
\end{document}